# Coupling Clinical Decision Support System with Computerized Prescriber Order Entry and their Dynamic Plugging in the Medical Workflow System

Lotfi BOUZGUENDA
Miracl laboratory/ University of Sfax/ISIM
Route de Tunis, km 10. BP 242, 3021 Sakeit Ezzit, Sfax-Tunisia
lotfi.bouzguenda@isimsf.rnu.tn

Manel TURKI
University of Monastir/ Faculty of pharmacy
Rue Ibn Sina, 5000, Monastir-Tunisia
Manel_Turki@yahoo.fr

*Abstract*. **This work deals with coupling Clinical Decision Support System (CDSS) with Computerized Prescriber Order Entry (CPOE) and their dynamic plugging in the medical Workflow Management System (WfMS). First, in this paper we argue some existing CDSS representative of the state of the art in order to emphasize their inability to deal with coupling with CPOE and medical WfMS. The multi-agent technology is at the basis of our proposition since (i) it provides natural abstractions to deal with distribution, heterogeneity and autonomy which are inherent to the previous systems (CDSS, CPOE and medical WfMS), and (ii) it introduces powerful concepts such as organizations, goals and roles useful to describe in details the coordination of the different components involved in these systems. In this paper, we also propose a Multi-Agent System (MAS) to support the coupling CDSS with CPOE. Finally, we show how we integrate the proposed MAS in the medical workflow management system which is also based on collaborating agents.**

## I. INTRODUCTION

The purpose of Clinical Decision Support System (or CDSS for short) is to assist health professionals with decision making tasks, as determining diagnosis or analysis of patient data [1].

In spite of the growing multiplicity of CDSS and their effectiveness certified in the decision making tasks at the time and the location of care, the state of the art of the existing CDSS ([2-6]) emphasizes four main challenges that require to be resolved. The first challenge is that the clinical data that must be entered is already contained elsewhere in a digital form in that hospital's system, and some CDSSs (alert system, drug-drug detection system, medicinal errors prevention system, etc.) are not able to access and exploit this information. The second challenge concerns the appropriate decision making by the health professionals to a patient. Indeed, the majority of existing CDSS exploit only the list of the actives prescripts of the patient, which is compared with a frozen source of knowledge fulfilled under form of mapping table. In others terms, they do not take into account the clinical context of the patient. A such CDSS must consider several information's such as patient's symptoms, medical history, family history and genetics. It needed to interact with others clinical data sources such as patient administrative data, physio-pathologic profile, biological analysis data, etc. The third challenge is that the CDSS is often not connected to Computerized Prescriber Order Entry (CPOE) which assists the doctor in preparing the prescripts.

The last challenge concerns the integration issue of the CDSS and CPOE in the medical workflow management system. In fact, often the clinical user must stop clinical process on the current system, switch to the CDSS or CPOE, and reenter data necessary into the CDSS that may already exist in another healthcare system.

An obvious example of clinical workflow is the patient care in an accident case. In this case, various health professionals such as the doctors, the nurses, the pharmacists, must cooperate, synchronize their intervention processes, share the access to the clinical data sources and use the CDSS, the CPOE without stop the clinical process to act in a coherent way and in order to give care to the patient.

Giving the previous observations, the problem being addressed in this paper can be resumed according to the following two questions: "how to design and develop a CDSS which considers the previous challenges? How do we integrate the CDSS and CPOE in the medical workflow management system?

The contribution of the paper is to provide a Multi-Agent System (MAS) to support the coupling CDSS with CPOE. Then, it explains how we integrate the proposed MAS in the medical Workflow management system.

Our approach is based on the following principles:

- The use of multi-agent technology, which provides relevant high-level features to design and implement CDSS, CPOE and clinical workflow management system. Indeed, Agent technology provides natural abstractions to deal with autonomy, distribution, heterogeneity and cooperation which are inherent to the previous systems,

- A coordination model based on organizational dimension. A good coordination model among clinical workflow components requires an organizational model which attributes roles to each component and constraints their interactions,

- An engineering perspective, which takes into account the existing standards. It defines an architecture which is compliant with the WfMC reference model [7] and the involved agents communicate among them using the FIPA-ACL standard [8]. This compliance allows a very large usability and interoperability of the solution.

The remainder of this paper is organized as follows. Section 2 exposes some related works regarding the design and the development of CDSS in order to underline their inability to deal with the challenges as mentioned previously. Section 3 presents our Multi-Agent System (MAS) for supporting the coupling CDSS with CPOE. This section first justifies the interest of use of multi-agent technology to design and develop the CDSS, CPOE and WfMS systems. It exposes then our proposed multi-agent system. Section 4 shows how we integrate dynamically the proposed MAS in the medical workflow management system and presents an organizational model based on message methodology [9] that structures the interactions between agents and highlights the coordination at macro level. Section 5 concludes the paper and gives some perspectives.

## II. RELATED WORKS

Regarding the design and development of CDSS, the use of agent technology is not new, and several works have been proposed in the literature ([2-6]).

[2] defends the interest of using agents to extend the medical expert systems. Indeed, according to the [2] the agents can resolve some issues by checking several conditions that could be ignored by humans and as a consequence the elimination of some mistakes from the physicians' decisions. More precisely, [2] proposes a new cooperative medical diagnosis system called "Contract Net Based Medical Diagnosis System". This latter owns two specific features namely the autonomy and the flexibility during the treatment of medical diagnosis problems. The proposed system in [2] does not support the interaction either with the CPOE or with the medical WfMS.

[3] proposes a CDSS called "SAPHIRE". The main purpose of this system is to support the definition, deploying and performing clinical guidelines to a patient. It is composed of a set of collaborating agents running in a heterogeneous distributed environment. SAPHIRE has two main advantages. Firstly, it proposes a specific agent called "EHR" agent" that is responsible to access and extract clinical data from the Electronic Healthcare. Secondly, it supports the interaction with several modules in the clinical workflow. The main limits of this work are the following: (i) the non possibility to couple with the CPOE and (ii) the use of agent only for the design of CDSS.

[4] provides a framework based on multi-agent system paradigm in order to build a comprehensive clinical decision applications aimed at various medical conditions. This work has not addressed the coupling issue with the healthcare systems such as CPOE, clinical workflow system, etc.

[5] proposes a multi-agent system called IMASC to assist physicians and other health professionals with decision making tasks. In spite of the system is powerful, it does not support the previous challenges, as mentioned in introduction, and namely the coupling with CPOE and the dynamic integration in the medical workflow management system.

[6] provides a prototype called MET3 which aims at data collection, diagnosis formulation and treatment planning. MET3 is based on multi-agent technology. It supports the interoperability issue since it runs on different platforms. It is also capable to interact which hospital information systems and particularly with an electronic patient record system via HL7 messages to provide realistic integration with existing healthcare systems.

Our work differs from the previous works on three points. First, it addresses the coupling CDSS with CPOE aspect which has not been addressed in the previous works. Second, the proposed architecture for medical Workflow management system is compliant with the Workflow Management Coalition reference model. Third, to the best of our knowledge, the dynamic plugging of CDSS and CPOE in the medical Workflow management system through the agent technology has not been addressed. Thus, we believe that our solution is currently unique in trying to take into account the agents to deal with coupling CDSS with CPOE and their dynamic integration in the medical WfMS.

## III. A MULTI-AGENT SYSTEM FOR COUPLING CDSS WITH CPOE

This section is devoted to the motivation for using multi-agent technology and presents our multi-agent system for coupling CDSS with CPOE.

### A. Motivation for using multi-agent technology

The multi-agent technology can help the design and the development of CDSS, CPOE and clinical workflow system thanks to the following high-level properties [10]:

- Autonomy of agents eases the cooperation since it avoids regular and direct interventions of the systems: the medical information system or clinical workflow management system can be agentified to support the cooperation and provides needed information to each health professional. Agents can also play the role of interface between the actors and the system (filtering and notification of events, providing relevant views of the whole system and its evolution).

- Natural abstractions to deal with cooperation. A lot of sophisticated protocols like Contract-Net Protocols

and Negotiation mechanisms are available and could be used to coordinate processes ([11] [12]). Agent technology also provides organizational concepts to abstract and structure a system as a computational community made of groups, roles and interaction.

- Pro-active and reactive attitudes of agents ease the control and enactment of clinical processes as well as reactions to events, and hence the synchronization of related activities. Being able to exhibit goal-directed behaviour, agents can take the initiative to select and engage cooperation with others actors.

- Social abilities of agents also ease the cooperation needed to enact complex clinical workflows and to provide an abstraction to high-level concepts like commitments, reputations and so on.

### B. *The proposed MAS for coupling CDSS with CPOE*

In order to support the coupling CDSS with CPOE, we propose a Multi-Agent System (MAS) where each agent plays a specific role or function and exploits one or several clinical data sources. More precisely, the architecture of our MAS is organized around five clinical data sources and ten specialized agent. Let us detail the role/function of each agent.

The communication between agents is assured thanks to the FIPA-ACL language [8].

The Information Collection Agent (ICA) is responsible to extract information's from two clinical data sources (patient administrative data and physio-pathologic profile) such as name, type, date of birth, service, weight, size, diagnosis, allergy, against indications, situations at risk and so on. It is to signal that these information's are managed by a medical information system and they can be also exploited by the others systems like CDSS and CPOE in our context. In this case, this agent answers to the two first challenges as mentioned in introduction i.e. we haven't need to enter these information's again.

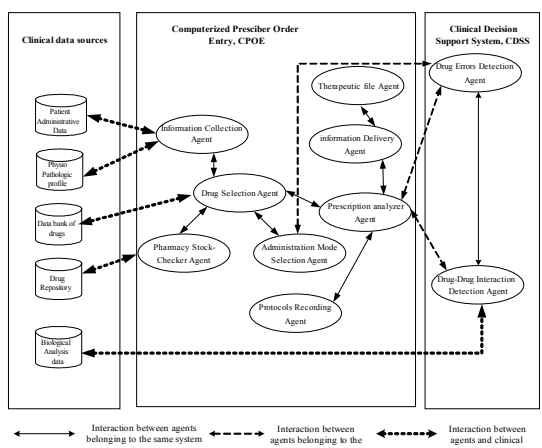

Figure 1. A MAS for Coupling CDSS with CPOE

The Drug Selection Agent (DSA) cooperates with the ICA and the Pharmacy Stock Checker Agent (PSCA) when selecting the drug form data bank of drugs. The role of the PSCA is to verify from the drug repository if the selected drug is available or in rupture of supply. In unavailability case, it recommends to substitute the drug by another generic. The DSA also interacts with the Administration Mode Selection Agent (AMSA) who provides relevant information's regarding administration mode, for each selected drug, such as unit of catch, dosage by default, and borders dosage by catch, a day and by kilogram of weight. To do this, it enters in connection with the Drug Errors Detection Agent.

The Prescription Analyzer Agent (PAA) triggers two agents to analyze and validate the current prescription: the Drug Errors Detection Agent (DEDA) and the Drug-Drug Interaction Detection Agent (DDIDA). More precisely, the Drug Errors Detection Agent (DEDA) interacts with the AMSA and DDIDA in order to verify if there is an error such as error of dose, error of choice (i.e. a drug not conforms to the reference), drug-drug interaction or error of administration mode and so on. In the literature, we distinguish two types of medicinal errors. A medicinal error is said potential if it was discerned before drug arrives up to the patient. It is said proved to be if it was discerned after the catch of drug by the patient. In our work, we support the first type of medicinal error in order to guarantee a well service to the patient.

The Drug-Drug Interaction Detection Agent (DDIDA) consists in proving if the association of medicaments is likely to cause undesirable effects, allergies, etc. To do this, it interacts with the Drug Errors Detection Agent and exploits the biological analysis data source.

The Protocols Recording Agent (PRA) stores all types of prescripts as protocols which will be then called by their title. As a consequence, the time of prescript becomes considerably reduced.

The Information Delivery Agent (IDA) is responsible to deliver the information to the concerned clinical user like the chemist, the nurse, etc.

The Therapeutic File Agent (TFA) preserves the validated prescription as archived files in order to be exploited in the future.

## IV. DYNAMIC PLUGGING OF CDSS AND CPOE IN THE MEDICAL WORKFLOW MANAGEMENT SYSTEM

The purpose of this section is (i) to rethink the reference model of a WfMS with agent technology in order to deal with medical processes and (ii) to show how we integrate the MAS that supports the coupling CDSS with COPE in the new architecture of WfMS.

### A. *The reference architecture of a WfMS*

According to this architecture, a WfMS includes a Workflow Enactment Service (WES) and supports the following interfaces (see figure 2) [7]:

- Interface 1 with Process Definition,
- Interface 2 with Workflow Client Applications,
- Interface 3 with Invoked Applications,

- Interface 4 with others WESs,
- Interface 5 with Administration and Monitoring.

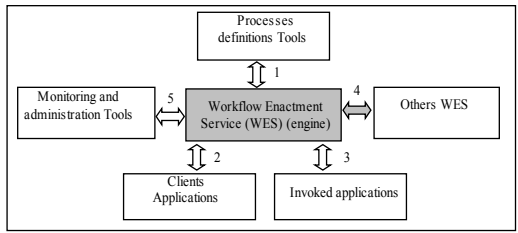

Figure 2. The reference model of a WfMS [7]

Two main components of this architecture are the WES and Interface 4. The aim of the WES, on which one imposes no constraints upon its internal structure, is to manage the execution of one or several instances of processes while the aim of Interface 4 is to connect WfMS together in order to share the execution of a workflow process between different WESs of different organizations.

However, this reference architecture is inadequate in the clinical context since the WES must not only execute clinical processes but also must mix different concurrent activities including clinical decision making, analyzing and validation of prescription, etc. In others words, the WES need to cooperate with CDSS and CPOE.

### B. Revisiting the WES with agents and dynamic plugging of CDSS and CPOE

Figure 3 explains how we have rethought the Workflow Enactment Service using agents. This architecture includes (i) as many Clinical Workflow agents as running clinical process instances, (ii) an Agent Manager responsible for these Clinical Workflow Agents, (iii) a Connector Agent that interacts with CDSS and CPOE specialized respectively in clinical decision making and prescription elaborating, and, (iv) a new interface, Interface 6, to support the communication between a Connector Agent and the proposed MAS.

Regarding clinical Workflow Agents, the idea is to implement each clinical process instance (stored in the Clinical Processes database, CP) as a software process, and to encapsulate this instance within an (pro-) active agent. Such a clinical Workflow Agent includes a workflow engine that, as and when the clinical process instance progresses, reads the CP schema definition (specified in XML schema), and triggers the action(s) to be done according to its current state. This clinical Workflow Agent supports interface 3 with the applications that are to be used to perform pieces of work associated to process' tasks.

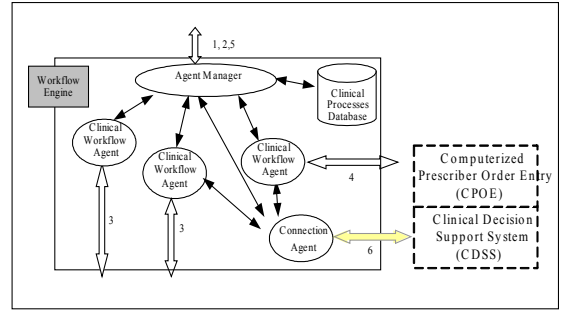

Figure 3. Dynamic plugging of the proposed MAS in the medical Workflow Management System

The Agent Manager controls and monitors the running of clinical Workflow Agents:

Upon a request for a new instance of a clinical process, the agent manager creates a new instance of the corresponding clinical process agent schema, initializes its parameters according to the context, and launches the running of its workflow engine.

It ensures the persistency of clinical Workflow Agents that execute long-term clinical processes extending to a long time in which task performances are interleaved with periods of inactivity.

It coordinates clinical Workflow Agents in their use of shared resources.

It assumes interfaces 1, 2 and 5 of the WfMS with the environment.

The role of the Connector Agent is to help clinical Workflow Agents to find the clinical information they need. More precisely, the Connector Agent interacts with CDSS specialized in clinical decision making. This requires defining a new interface, Interface 6 that supports the communication between a Connector Agent, the CDSS and the COPE. Interface 4 of the reference architecture cannot be used for such a communication since it only supports the execution of a clinical process between different workflow engines.

In our proposition, we consider coordination as a specific component when designing and implementing a clinical Workflow, and consequently, we separate coordination activities from clinical processes execution. That is why we introduced mediation infrastructure, gathering the CDSS and COPE. This infrastructure of course independent of the WES, and is dynamically plugged to it only when it is necessary.

The standard FIPA-ACL (Foundations of Intelligent Physical Agents-Agent Communication Language [8]), is used to support the interaction, through message exchange, between the agents inside the architecture of WfMS. FIPA-ACL offers a convenient set of performatives for supporting the cooperation between agents (e.g. inform, ask, propose, agree, cfp,...). Moreover, FPA-ACL supports exchange messages between heterogeneous agents since (i) the language used to specify the message is free and (ii) a message can refer to ontology. This is very interesting in the context of inter organization clinical processes since ontology can be used to solve semantic interoperability problems.

**C.** *An organizational model based on message methodology for ruling the interaction between the clinical workflow components*

Our organizational model (see figure 4) is organized around the following components:

- Three organizations represented by triangles:
- Ten agents represented by circles :
- Nine tasks represented by polygons:
- Three roles represented by moon:
- and three goals represented by double circles:

For clarity reason of the organizational model we give only the tasks, goals and roles for three agents (Agent Manager and Drug-Drug Interaction Detection Agent and Prescription Analyzer Agent.

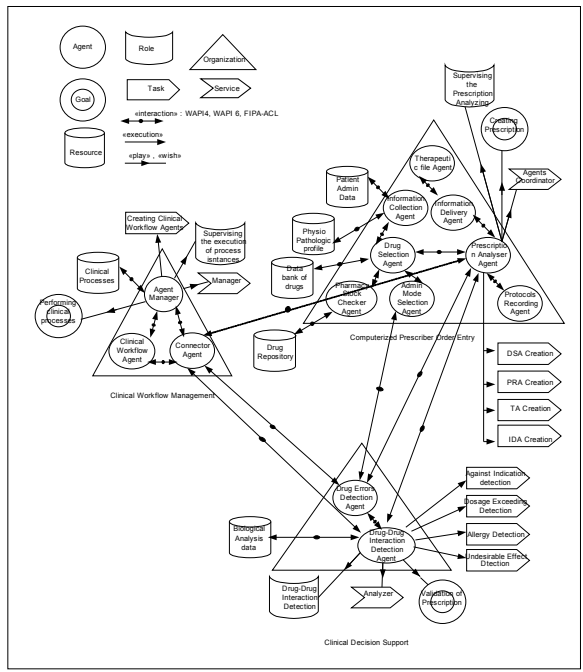

Figure 4. An organizational model for structuring the communication between CDSS, CPOE and clinical WfMS

Let us detail how each organization operates.

The Clinical Workflow Management Organization enables one or several clinical workflow agents to interact with their connector to which make a clinical decision or elaborate a validated prescription.

The Computerized Prescriber Order Entry Organization enables the Prescription Analyzer Agent to interact at the first step with the drug selecting Agent.

This latter interacts in its turn with the information collection agent, pharmacy stock checker Agent and Administration mode selection agent. At the second time, the Prescription Analyzer Agent connects to the Drug-Drug interaction Detection Agent and Drug Errors Detection Agent to analyze the current prescription. After the validation, it offers the following services thanks to the three agents: Information delivery, protocols recording and therapeutic file creation.

The Clinical Decision Support Organization enables the drug errors detection and drug-drug interaction detection agents to enter in connection with the Connector Agent or the Prescription Analyzer Agent in order to give respectively help to the clinical user or assist the doctor to validate his prescription. This organization involves only those two agents.

## V. CONCLUSION

This paper has dealt with coupling CDSS with CPOE and their dynamic plugging in the medical Workflow Management System. More precisely, it has presented in the one hand, a multi-agent system allowing the coupling CDSS with CPOE. In the other hand, it has presented an architecture of clinical WfMS which revisits the reference model defined by the WfMC in terms of collaborating agents. Some of these agents implement clinical processes, while others are dedicated to the CDSS and CPOE.

The idea defended in this paper is that the agent technology is appropriate to face the coupling CDSS with CPOE and their dynamic integration in the clinical WfMS. Several reasons motivate this idea.

Firstly, the agent technology is appropriate to model clinical processes since it provides natural abstractions to deal with distribution, heterogeneity and autonomy that are inherent to clinical processes.

Secondly, the organizational dimension which is inherent to multi-agent systems [9], is fundamental to highlight the coordination of the different components involved in the three systems (CDSS, CPOE and clinical WfMS) under consideration by clearly separating the macro-level (coordination) from the micro-level (agent). Thus, the internal architecture of the agents may be thought and implemented independently from their coordination.

As future work, we plan (i) to study in depth the behavior of all agents involved in the CDSS, CPOE and WfMS systems, (ii) to formalize and build the knowledge's used by the agents in order to resolve the semantic conflicts, (iii) to implement these agents using the Jade [13] (Java Agent Development Framework) platform that offers the possibility to simulate the execution of these agents in autonomic way and in distributed mode and (iv) to evaluate the proposed tool according to the three key criteria: scalability, openness and efficiency.